\documentclass[12pt]{iopart}

\usepackage{latexsym}
\usepackage{graphicx} 
\usepackage{psfig}

\newcommand{\newtxt}{}

\begin{document}

\title[Universality of free fall for strongly self-gravitating bodies]{Tests of 
the universality of free fall for strongly self-gravitating bodies with radio 
pulsars}

\author{Paulo C.~C.~Freire, Michael Kramer and Norbert Wex}

\address{Max-Planck-Institut f\"ur Radioastronomie, 
	Auf dem H\"{u}gel 69, D-53121 Bonn, Germany}
\ead{pfreire@mpifr-bonn.mpg.de}

\begin{abstract}
In this paper, we review tests of the strong equivalence principle (SEP) derived 
from pulsar-white dwarf data. The extreme difference in binding energy between both 
components and the precise measurement of the orbital motion provided by pulsar 
timing allow {\newtxt the only current precision SEP tests for strongly 
self-gravitating 
bodies}. We start by highlighting why such tests are conceptually important. We 
then review previous work where limits on SEP violation are obtained with an 
ensemble of wide binary systems with small eccentricity orbits. {\newtxt Then we 
propose a new SEP violation} test based on the measurement of the variation of 
the orbital eccentricity ($\dot{e}$). {\newtxt This new method has the following 
advantages: a) {\newtxt unlike previous methods} it is not based on 
probabilistic considerations, b) it can make a direct detection of SEP 
violation, c) the measurement of $\dot{e}$ is not contaminated by any 
known external effects, which implies that this
SEP test is only restricted by the measurement precision of 
$\dot{e}$}. In the final part of the review, we conceptually compare the SEP 
test with the test for dipolar radiation damping, a phenomenon closely related 
to SEP violation, and speculate on future prospects by new types of tests in 
globular clusters and future triple systems.
\end{abstract}


\section{Introduction}

The Strong Equivalence Principle (SEP) extends the Weak Equivalence Principle 
(WEP) to the universality of free fall (UFF) of self-gravitating bodies.
General Relativity (GR) assumes the validity of WEP and SEP, i.e.~in GR the world line 
of a body is independent of its chemical composition and gravitational binding 
energy. Therefore, a detection of SEP violation would falsify GR.
On the other hand, alternative theories of gravity generally violate 
SEP. This is also the case for most alternative {\em metric} theories,
such as the Jordan-Fierz-Brans-Dicke (JFBD) theory \cite{jor59,fie56,bd61}
or the more general scalar-tensor theories of gravity \cite{de92},
which satisfy WEP as a consequence of their postulate of universal
coupling between matter and gravity (\cite{wil93}, see also review by
T.~Damour in this issue)\footnote{In fact, Willy Scherrer was the first one who
had proposed a scalar-tensor theory of gravitation (see \cite{goe12} for
details on the history of scalar-tensor theories).}.

SEP violation could have observable consequences even in the weak, 
quasi-stationary gravitational fields in our Solar System \cite{nor68a,nor68b}, 
in particular a ``polarization" of the Earth-Moon system by the external Solar 
field \cite{nor68c}. Detecting such a polarization is one of the main 
motivations of the Lunar-Laser-Ranging (LLR) experiment, which is described in 
other articles in this issue.

In view of the smallness of self-gravity of Solar System bodies, the LLR
experiment says nothing about strong-field aspects of gravity, like the 
gravitational properties of extremely compact objects like neutron stars or
black holes. In the presence of such bodies, alternative theories of gravity 
that pass Solar System tests can still produce various observable phenomena 
not predicted by GR, 
among these the {\em radiation of dipolar gravitational waves} and
{\em strong-field SEP violation} (Section~\ref{sec:SEP_theory}). The former has not yet been detected in 
the most sensitive binary pulsar experiments \cite{ksm+06,fwe+12}, which given 
the current experimental precision already provide extremely stringent tests of
alternative theories of gravity. In the near future the detailed study of
the gravitational wave signal emitted by mergers of compact objects will also 
provide independent tests of the radiative properties of gravity (e.g., 
\cite{wil94}).

For strong-field SEP violation, the best current limits 
come from millisecond pulsar (MSP) - white dwarf (WD) systems with wide orbits.
If there is a violation of UFF by neutron stars, then the gravitational
field of the Milky Way would polarize the binary orbit \cite{ds91}.
Such tests of the UFF of strongly 
self-gravitating masses are the subject of this work. In comparison with the LLR 
tests, they have two disadvantages, one of them being the much weaker polarizing 
external field ($|{\bf{g}}| \sim 2 \times 10^{-10}\rm\, m\,s^{-2}$ compared with 
the acceleration of
the Solar gravitational field at the Earth, $\sim 6 \times 10^{-3}\rm\, m
\,s^{-2}$) and the precision of the ranging,
which is of the order of $10\,$m
for the best pulsar experiments ($10^{-2}\,$m for LLR). 
This is almost completely compensated by the
gravitational binding energy of neutron stars, $E_{\rm grav}$, which is
a large fraction of the total inertial mass-energy:
$\varepsilon_{\rm grav} = E_{\rm grav} / M_{\rm I}c^2 \sim - 0.15$; this
is more than eight orders of magnitude larger
than Earth's ($\varepsilon_{\rm grav, \oplus} \sim -5 \times 10^{-10}$).
This results in experiments with comparable limits on SEP violation,
which are nonetheless complementary since they probe different regimes of 
binding energy.

In Section~\ref{sec:SEP_theory} we summarize some of the theoretical 
foundations of SEP violation for strongly self-gravitating bodies. In
Section~\ref{sec:DS_test} we review the Damour-Sch\"afer test, which yields the 
best current limits on SEP violation by strongly self-gravitating bodies,
if applied to a whole population of small-eccentricity systems.

In Section~\ref{sec:Detecting} we suggest new pulsar timing experiments that 
avoid the probabilistic considerations of present tests and have the potential 
to detect SEP violation; these attempt to directly measure its effects,
in particular the variation of the orbital eccentricity, $\dot{e}$.
It is shown that this 
measurement is not contaminated by external effects and, because of this, the 
limits on SEP violation are only restricted by the precision of the measurement
of $\dot{e}$. Our simulations 
show that such direct tests will very soon surpass the best current limits on 
SEP violation for strongly gravitating bodies.

Section~\ref{sec:complementarity} provides a short 
discussion on the complementarity of SEP tests and tests for dipolar radiation 
damping in constraining alternative theories of gravity.
Finally, in Section~\ref{sec:future} we briefly discuss the possibilities
of the SEP test proposed Section~\ref{sec:Detecting} when the test
binaries are accelerating in a fields much stronger
than that of the Galaxy, like that of a globular cluster or an additional
outer companion. Such systems would likely offer very significant gains in
the power of SEP tests, essentially providing an experiment with fields
and velocities very similar to those present LLR experiment, but with
``proof masses" with a $\sim 10^8$ larger binding energy - essentially
a combination of the best features of both types of experiments. 


\section{SEP violation and strongly self-gravitating bodies}
\label{sec:SEP_theory}


\subsection{Beyond the weak-field approximation}

In general, an alternative theory of gravity is expected to violate the strong
equivalence principle, in the sense that the ratio between the gravitational
mass $M_{\rm G}$ of a self-gravitating body to its inertial mass $M_{\rm I}$ 
will admit an expansion of the type 
\begin{equation}\label{eq:mgmi}
  \frac{M_{\rm G}}{M_{\rm I}} 
    \equiv 1 + \Delta 
    =      1 + \eta_1 \, \varepsilon_{\rm grav} + 
               \eta_2 \, \varepsilon_{\rm grav}^2 + \dots
\end{equation}
At the first 
post-Newtonian level $\eta_1$ parametrizes the weak-field violation of SEP 
(Nordtvedt effect). In the Parameterized Post-Newtonian (PPN)
framework {\newtxt this {\it Nordtvedt parameter}} is 
given as a combination of different PPN parameters (see \cite{wil93} for details 
on the PPN formalism): 
\begin{equation}
  \eta_1 \equiv \eta_N = 4\beta - \gamma - 3 - \frac{10}{3}\xi
                                - \alpha_1 + \frac{2}{3} \alpha_2
                                - \frac{2}{3}\zeta_1 - \frac{1}{3}\zeta_2 \;.
\end{equation}
{\newtxt The parameter $\eta_N$} is well constrained by lunar laser ranging (LLR) 
experiments in the Solar System (see contributions on LLR in this review). In
view of the fact that $\varepsilon_{\rm  grav, \oplus} \sim -5 \times 10^{-10}$,
$\varepsilon_{\rm  grav, Moon} \sim -2 \times 10^{-11}$ and 
$\varepsilon_{\rm  grav, \odot} \sim -10^{-6}$, it is clear that higher order 
deviations from SEP cannot be tested in LLR experiments, or any other experiment 
in the Solar System, in the foreseeable future. To test a violation of SEP that 
might occur beyond the weak-field regime, one needs strongly self-gravitating 
bodies. Presently, in nature the best objects for such tests are neutron stars 
observed as active radio pulsars ($\varepsilon_{\rm grav} \sim 
-0.15$).\footnote{For gravity theories that predict a no-hair theorem for black 
holes, neutron stars are in fact the most compact objects suitable for
strong-field SEP violation tests (see e.g.~\cite{de98,ppdm08,abwz12}).}

Since beyond the weak field approximation there is no general PPN formalism 
available, discussions of gravity tests in this regime are done in various 
theory specific frameworks. An example for a very detailed investigation of 
higher order/strong-fied deviations from GR, within the family of 
{\newtxt (well-defined)} scalar-tensor 
theories of gravity, are the frameworks developed by Damour and 
Esposito-Far{\`e}se. Specifically,
\begin{description}
\item[A)] The 4-parameter framework $T_0(\gamma, \beta; \epsilon, \zeta)$ of 
          \cite{de96a}, which defines the second post-Newtonian extension of 
          the original (Eddington) PPN framework $T_0(\gamma, \beta)$. In this 
          framework 
          \begin{equation}
            \Delta_A \simeq -\frac{1}{2}(4\beta - \gamma - 3) c_A +
            \left(\frac{\epsilon}{2} + \zeta\right) b_A \;,
          \end{equation}
          where the compactness $c_A \approx -2\varepsilon_{{\rm grav},A}$ 
          and $b_A \approx c_A^2$.
\item[B)] The 2-parameter class of bi-scalar-tensor theories $T_2(\beta', 
          \beta'')$ introduced in \cite{de92}. Within this framework, one has 
          no Nordtvedt effect in the Solar System, since $\eta_{\rm N} = 0$.
          On the second post-Newtonian level one has $\epsilon = \beta'$, 
          $\zeta = 0$. And $\beta''$ parametrizes contributions beyond the 
          second post-Newtonian level.
\item[C)] The 2-parameter class of mono-scalar-tensor theories $T_1(\alpha_0, 
          \beta_0)$ of \cite{de93,de96b}, which for certain values of $\beta_0$
          exhibits significant strong field deviations from GR, and a 
          corresponding violation of SEP for neutron stars.
\end{description}
For details we refer the reader to \cite{dam07} and references therein. {\newtxt In 
the following we will have a closer look at case C, as this class of 
scalar-tensor theories of gravity illustrates quite impressively a violation of 
SEP that can only be measured with strongly self-gravitating bodies. We would 
like to note, however, that this particular case of scalar-tensor theories of 
gravity resembles just one example of how non-linearities in the gravitational 
interaction could drive gravity away from GR in the strong fields of compact 
masses and lead to a strong-field violation of SEP.}


\subsection{Strong field effects and the violation of SEP.}

Damour and Esposito-Far{\`e}se \cite{de93} found that scalar-tensor theories, 
which pass the weak-field tests in the Solar System, could still exhibit large, 
strong-field-induced deviations in systems involving neutron stars 
(``spontaneous scalarization''). This has been studied extensively in a two-
parameter space of theories, $T_1(\alpha_0,\beta_0)$, defined by the coupling 
function which is a quadratic polynomial in the scalar field $\varphi$: 
$a(\varphi) = \alpha_0 (\varphi - \varphi_0) + \beta_0(\varphi - \varphi_0)^2/2$ 
\cite{de93,de96b,de98}. {\newtxt The parameter $\alpha_0$ defines the linear 
matter-scalar coupling constant and $\beta_0$ the quadratic coupling of matter 
to two scalar particles, while higher-order vertices are neglected. In this 
sense, this is a natural extension of JFBD gravity.}

In presence of such non-perturbative strong-field deviations away from GR, we 
can have a situation where the effective coupling strength of the neutron star, 
$\alpha_A$, is of order unity, even if the scalar-matter coupling, $\alpha_0$, 
is unobservably small in the Solar System\footnote{\newtxt The quantity $\alpha_A 
\equiv \partial \ln M_A / \partial \varphi_0$ measures the effective strength of 
the coupling between a self-gravitating body $A$, with total mass $M_A$, and the 
scalar field $\varphi$. It is equivalent to the negative ratio of total scalar 
charge to total mass. For a weakly self-gravitating body $\alpha_A \simeq 
\alpha_0$.}. Such an effect leads to a violation of SEP that requires test 
systems which contain a neutron star. 

The structure dependence of the effective gravitational constant $G_{AB}$, has 
the consequence that the pulsar does not fall in the same way as its companion  
in the gravitational field of a third body, which in our case is the Galaxy. 
One finds for a pulsar with a weakly self-gravitating companion, since $\alpha_0 
\ll 1$, that \cite{de92}
\begin{equation} \label{eq:DeltaTS}
  \Delta_{\rm p} - \Delta_{\rm c}  
  \simeq \alpha_0 (\alpha_{\rm p} - \alpha_{\rm c}) 
  \simeq \alpha_0 (\alpha_{\rm p} - \alpha_0) \;.
\end{equation}
While $|\alpha_0| < 0.003$ by the Cassini experiment \cite{bit03}, 
$\alpha_{\rm p}$ 
can be of order unity for neutron stars, as outlined above. Although the effect
is greatly suppressed by a small factor $\alpha_0$, it could in the presence of
non-perturbative strong-field deviations be many orders of magnitude stronger 
than in the Solar System (factor $\alpha_{\rm p} / \alpha_0$), and therefore 
still become visible in binary pulsar timing experiments.


\subsection{SEP violation and orbital motion of binary pulsars}

In the previous section we have seen that one could expect violations of SEP in 
gravity regimes where the Solar System tests are insensitive to, and therefore 
one has to utilize test systems that contain strongly self-gravitating bodies. 
Currently, binary pulsars are the best probes for testing such kind of strong 
field 
gravity effects \cite{dam07}. Before we discuss in details the SEP tests that 
can be conducted in timing binary pulsars, we need to outline the orbital 
dynamics of a binary system that falls freely in the gravitational field of our 
Galaxy in the presence of a SEP violation.

In case of a violation of SEP the equations for the relative motion 
${\bf{R}}(t)$ are given by
\begin{equation}
  \ddot{\bf{R}} = -{\cal G}M\frac{\bf{R}}{R^3} + 
                  {\bf{A}}_{\rm PN} + {\bf{A}}_\Delta \;
\end{equation}
{\newtxt where $M$ is the total mass of the binary and} 
${\cal G} \equiv G_{\rm pc}$ is the effective gravitational constant 
between the two bodies \cite{ds91}. The post-Newtonian contributions are denoted 
by ${\bf{A}}_{\rm PN}$ and the additional acceleration caused by SEP violation, 
${\bf{A}}_\Delta$, is given {\newtxt to leading order} by\footnote{{\newtxt As 
the gravitational field of the Galaxy at the location of the binary pulsars is 
weak, and since most of the mass of the Galaxy is made of non 
strongly-self-gravitating bodies, any higher order contributions to 
${\bf{A}}_\Delta$ are negligible.}}
\begin{equation}
  {\bf{A}}_\Delta = (\Delta_{\rm p} - \Delta_{\rm c}) \, {\bf{g}} \;, 
\end{equation}
where $\Delta_i$ is defined by the ratio between ``gravitational'' and 
``inertial'' mass as given in eq.~(\ref{eq:mgmi}), and ${\bf{g}}$ is the external  
acceleration caused by the gravitational field of the Galaxy. The secular 
changes to the binary system with orbital frequency $n_{\rm b}$ and eccentricity 
$e$ caused by a violation of SEP are \cite{ds91}
\begin{equation}\label{eq:dtsep}
  \langle dn_{\rm b}/dt \rangle = 0 \;, \quad
  \langle d{\bf{e}}/dt \rangle = {\bf{f}} \times {\bf{l}} \,+\, 
    \dot{\omega}_{\rm PN}\,\hat{\bf{k}}\times{\bf{e}} \;, \quad 
  \langle d{\bf{l}}/dt \rangle = {\bf{f}} \times {\bf{e}} \;,
\end{equation}
where 
\begin{equation}\label{eq:elfdefs}
  {\bf{e}} \equiv e \, \hat{\bf{a}} \;, \quad
  {\bf{l}} \equiv \sqrt{1 - e^2} \, \hat{\bf{k}} \;, \quad
  {\bf{f}} \equiv \frac{3}{2{\cal V}_O}\,
                 (\Delta_{\rm p}-\Delta_{\rm c})\,{\bf{g}} \;, 
\end{equation}
and ${\cal V}_O \equiv ({\cal G}Mn_{\rm b})^{1/3}$ is a measure for the relative 
{\newtxt orbital} velocity between the pulsar and its companion. The unit vector 
$\hat{\bf{a}}$ 
points towards periastron and the unit vector $\hat{\bf{k}}$ is {\newtxt parallel to 
the orbital angular momentum}. The post-Newtonian precession of periastron is 
given by
\begin{equation}
  \dot{\omega}_{\rm PN} = 
    3{\cal F}\,\frac{({\cal V}_O/c)^2}{1 - e^2}\,n_{\rm b} \;,
\end{equation}
where ${\cal F}$ is a theory-dependent factor (${\cal F} = 1$ in GR). As a 
result of eqs.~(\ref{eq:dtsep}), a violation of the UFF would lead to a change 
in the observed orbital eccentricity
\begin{equation}\label{eq:edotSEP}
  \dot{e} = \langle d{\bf{e}}/dt \rangle \cdot \hat{\bf{a}} 
          = \sqrt{1 - e^2} \, (\hat{\bf{b}} \cdot {\bf{f}}) \;, 
            \qquad (\hat{\bf{b}} \equiv \hat{\bf{k}} \times \hat{\bf{a}})
\end{equation}
and a change in the angle $i$ between the line-of-sight direction to the pulsar, $\hat{\bf{K}}_0$, and $\hat{\bf{k}}$
\begin{equation}
\label{eq:xdotSEP}
  \frac{d\cos i}{dt} = \frac{d}{dt} \left(
    \frac{ \hat{\bf{K}}_0 \cdot {\bf{l}}}{\sqrt{1 - e^2}} \right)
    =  \frac{e}{\sqrt{1 - e^2}}(\hat{\bf{K}}_0 \cdot \hat{\bf{b}})
       (\hat{\bf{k}} \cdot {\bf{f}}) \;.
\end{equation}
In binary pulsar timing experiments this change in $i$ becomes apparent as a 
change in the timing parameter $x$, which is the projected semi-major axis of 
the pulsar orbit given by $x = a_p \sin i/c$.

\begin{figure}
  \centerline{\includegraphics[height=5.0cm]{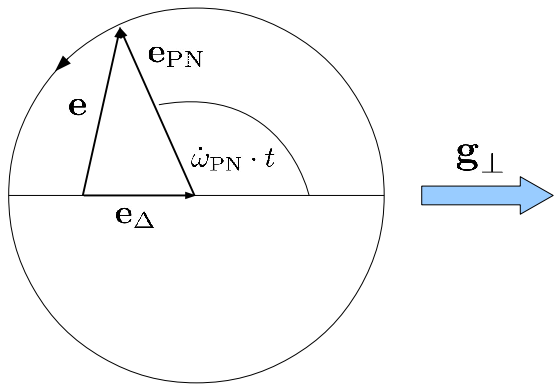}}
  \caption{Time evolution of the observed orbital eccentricity vector ${\bf{e}} = 
  {\bf{e}}_{\rm PN} + {\bf{e}}_\Delta$ in a small eccentricity binary. The
  vector ${\bf{g}}_{\perp}$ represents the projection of the external
  acceleration in the orbital plane.
  \label{fig:et}}
\end{figure}

As shown in \cite{ds91}, for very small eccentricities the equations of motion 
(\ref{eq:dtsep}) essentially decouple. As a consequence, the orbital plane 
remains fixed and the evolution of the eccentricity can be written as
\begin{equation}
  {\bf{e}}(t) = {\bf{e}}_{\rm PN}(t) + {\bf{e}}_\Delta \;, \quad
  {\bf{e}}_\Delta \equiv {\bf{f}}_\perp / \dot{\omega}_{\rm PN} \;.               
\end{equation}
The vector ${\bf{e}}_{\rm PN}(t)$ has a fixed length and is turning in the 
orbital plane with angular velocity $\dot{\omega}_{\rm PN}$. The polarization of 
the orbit due to the violation of SEP is represented by the constant 
eccentricity vector ${\bf{e}}_\Delta$, which points into the direction of the 
projection of ${\bf{f}}$ into the orbital plane (denoted by ${\bf{f}}_\perp$). 
Figure~\ref{fig:et} illustrates the time evolution of the orbital eccentricity. 
Thus, in the presence of a SEP violation the observed eccentricity oscillates 
between a minimum ($|e_{\rm PN} - e_\Delta|$) and maximum ($|e_{\rm PN} + e_
\Delta|$) eccentricity with a period of $2\pi/\dot{\omega}_{\rm PN}$.

\section{The Damour-Sch\"afer test}
\label{sec:DS_test}

In 1991, when Damour and Sch\"afer first wrote their paper describing the orbital 
dynamics of a binary pulsar under the influence of SEP violation, only four 
binary pulsars were known in the Galactic disk. Two of these 
(PSR~B1913+16 and PSR~B1957+20) were clearly inadequate for that test 
--- not only because of the compactness of their orbits, but also because one of 
them is a double neutron star system that lacks the required amount of asymmetry 
in the binding energy, necessary for a stringent test of SEP violation
[the term $(\alpha_p - \alpha_c)$ in eq.~(\ref{eq:DeltaTS})].
The remaining systems were PSR~B1855+09 \cite{rt91} and PSR~B1953+29 
\cite{bbf83}. The latter was already deemed to be the best probe for the 
detection of SEP violation effects because of its orbital period of 117 days, by 
far the largest then known. Nevertheless, even for this system, the orbital 
effects predicted for allowed levels of SEP violation --- in particular, the 
changing eccentricity $\dot{e}$ --- were too small for detection given the
timing precision and baseline of that system in 1991. Another way of
saying this is that at that time no interesting limits on SEP violation could be 
derived from existing constraints of $\dot{e}$. It is for this reason that 
Damour and Sch\"afer proposed {\newtxt a novel}, indirect statistical SEP test, by 
assuming that SEP violation is responsible for a maximal part of the observed 
small orbital eccentricities.


\subsection{Small-eccentricity binary pulsars and SEP}

The basic idea presented by Damour and Sch\"afer has been already
described above, namely that the violation of the SEP introduces a
polarization of the orbit of binary pulsars that is best represented
by a vector addition where the observed eccentricity vector ${\bf{e}}(t)$
lies on a circle (see Fig.~\ref{fig:et}). In the extreme case, however, the 
vector addition leads to a near or even complete cancellation of the 
eccentricity vector and a SEP violation is therefore not detectable. 
Unfortunately, the intrinsic vector ${\bf{e}}_{\rm PN}$, and therefore also its 
orientation relative to the SEP component vector ${\bf{e}}_\Delta$, is unknown.  
However, for a sufficient age of the system one can
assume that the relativistic precession of the orbit will have caused
the eccentricity vector to have made many turns since the system's
birth, thereby effectively randomizing the relative orientation
$\theta \equiv \dot{\omega}_{\rm PN} t$. In
fact, the angular velocity of the periastron advance, 
$\dot{\omega}_{\rm PN}$, should be appreciably larger than the angular velocity  
of the rotation of the Galaxy with which {\bf{g}} rotates in the reference frame 
of the binary system. As a result, the projection of the Galactic acceleration
vector onto the orbit can be considered constant, and statistical arguments 
based on the exclusion of small $\theta$, i.e.~possible near cancellation, can 
be applied. In the ideal case, the masses of the system components, the
inclination of the orbit and the distance to the pulsar are known.
The angle describing the orientation of the orbit around the line-of-sight, 
$\Omega$, is generally not observable and has to be treated as an independent 
random variable uniformly distributed between 0 and $2\pi$. 

As the treatment above has been derived for small-eccentricity, long-orbital
period pulsars, the figure-of-merit for suitable systems considered by
Damour and Sch\"afer was large values of $P_{\rm b}^2/e$ --- {\newtxt the large
values of $P_b$ decrease the relative orbital velocities ${\cal V}_0$,
which increases the amplitude of ${\bf{f}}$ (eq.~\ref{eq:elfdefs})
and the predicted change of eccentricity (eq.~\ref{eq:dtsep}); the small
value of $e$ implies that, in a statistical sense the orbit has been
little changed by such SEP violation effects.

They applied their method to the two systems
with (by far) the best $P_b^2/e$ at the time},
PSR B1855+09 and PSR B1953+29. While the measurement of a
Shapiro delay \cite{sha64,fw10} for PSR B1855+09 \cite{rt91,ktr94} were 
available, providing {\newtxt constraints on $\sin i$ and the component masses},
evolutionary arguments were used to constrain the parameters for
PSR~B1953+29. Overall, Damour and Sch\"afer derived 90\% confidence level 
limits of $|\Delta_{\rm p}|<5.6\times 10^{-2}$ and $|\Delta_{\rm p}|<1.1\times 
10^{-2}$, respectively.

Six years later, \cite{wex97} presented an updated analysis, applying this 
Damour-Sch\"afer test to eight systems with large values 
for $P_{\rm b}^2/e$, this excluded SEP violation at a
level of $5\times 10^{-3}$ with more than 95\% confidence.
    

\subsection{A population of small-eccentricity binary pulsars}
\label{sec:population}

The analysis presented in \cite{wex97} has a caveat, by selecting only those 
systems with the best figure-of-merit $P_{\rm b}^2/e$. By this, one introduces a 
selection bias, as it is possible that on the one hand a significant 
eccentricity (which would reduce the figure-of-merit and a possible weight in 
the analysis) is actually the result of an SEP violation, and on the other hand 
the small eccentricities selected are those where by chance $\theta$ is small. 
(This was pointed out to one of us (NW) by Kenneth Nordtvedt.)

In order to take this into account, \cite{wex00} presented an updated analysis 
that included all relevant small-eccentricity binary pulsars at that time. Based 
on extensive Monte-Carlo simulations, a large set of simulated (cumulative) 
distributions of eccentricities was compared to the distribution of 
eccentricities observed in the population of small-eccentricity binary pulsars. 
In this analysis, unknown angles like $\theta$ and $\Omega$ were distributed 
uniformly between $0$ and $2\pi$. Based on a Kuiper's test like criterion, the 
number of simulated distributions which are in agreement with the observed 
distribution was determined. As a result, \cite{wex00} found a 95\%-confidence 
limit of $|\Delta_{\rm p}| < 9\times 10^{-3}$. This limit is weaker than the one 
in \cite{wex97}, but certainly more representative by being able to consistently 
include also systems with seemingly worse figure-of-merits.  

In the same spirit, \cite{sfl+05} and later \cite{gsf+11} presented
an updated analysis, using even larger
samples of pulsar-WD systems, i.e.~the population of {\em all} known 
systems that are thought to have evolved with similar extended accretion 
periods. For deriving the median-likelihood value of $|\Delta_{\rm p}|$ for each 
pulsar, they use a Bayesian analysis marginalising over the parameters with 
similar assumptions to those in \cite{ds91,wex97}.
In systems like PSRs 0437$-$4715 and J1713+0747 where the orientation
of the orbit can be measured  \cite{vbb+01,sns+05}, posterior probability
density functions were derived appropriately.
{\newtxt From all available information they obtained
$|\Delta_{\rm p}| < 5.6\times 10^{-3}$ (95\% C.L.).
The best present} limit was obtained by 
\cite{gsf+11} using 27 binary systems (including additional astrometric and 
mass information), $|\Delta_{\rm p}| < 4.6\times 10^{-3}$ (95\% C.L.).


\section{Detecting SEP violation directly}
\label{sec:Detecting}

Many new advances in pulsar astronomy have occurred in the two decades elapsed 
since the publication of \cite{ds91}. Three of them are especially important
in this context:
\begin{description}
\item[A)] Continuing pulsar surveys have discovered many more binary systems, 
  including several with significantly wider orbits than PSR~B1953+29, i.e., 
  systems more suitable for the detection of SEP violation (Section~\ref{sec:DS_test}),
\item[B)] Advances in the sensitivity and bandwidth of radio receivers and   
  improvements in the time resolution of instrumentation mean that we
  can time the binary pulsars in these systems much more precisely than 
  possible before, 
\item[C)] The sheer amount of elapsed time provides much larger timing 
  baselines.
\end{description}
All of these developments increase the sensitivity to the direct effects of SEP 
violation, in particular $\dot{e}$ (Section~\ref{sec:figure_of_merit}), to the 
point that it should now be possible to derive even more stringent limits on SEP 
violation from them (Section~\ref{sec:present_limits}). This has several 
advantages over the statistical method described in Section~\ref{sec:DS_test}:
\begin{enumerate}
{\newtxt 
\item We can actually {\em detect} a potential SEP violation by measuring 
  a non-zero $\dot{e}$, particularly if the same phenomenon is observed for 
  several binary MSPs. With the statistical method reviewed in 
  Section~\ref{sec:DS_test} we can only estimate upper limits for the effect.
\item For wide pulsar-WD systems, only this hypothetical SEP violation
  can cause a measurable $\dot{e}$: orbital 
  circularization due to emission of gravitational waves
  or aberration effects cause a change in eccentricity that is
  many orders of magnitude below 
  the current experimental precision on $\dot{e}$; furthermore there are no 
  effects due to
  mass loss in the system or its motion (Section~\ref{sec:clean}). In this
  sense, this is a clean test of the validity of GR.
\item  
  The clean nature of the $\dot{e}$ test implies that
  the limit on SEP violation will improve at the same rate as the precision of 
  the
  measurement of $\dot{e}$. Thus, if the latter improves continuously with time
  (Section~\ref{sec:figure_of_merit}), the same will happen with limits on
  SEP violation}.
\item Since the test can be done with a single binary system, we do not need to 
  assume that $\Delta_{\rm p}$ is the same for all systems, as {\newtxt assumed in 
  the statistical test of Section~\ref{sec:population}}. Indeed, as pointed out 
  in \cite{dam07}, alternative 
  theories of gravity predict that $\Delta_{\rm p}$
  is a function of the pulsar mass. Therefore, a rigorous analysis requires an 
  accurate knowledge of the masses of the pulsars, which for many of the 
  pulsar-WD systems {\newtxt are} not available. In view of strong-field effects 
  like ``spontaneous scalarization'', the combination of binary pulsars in a 
  generic SEP test could even be rendered meaningless.
\item We do not need to restrict our sample to systems with small eccentricities  
  (Section~\ref{sec:1903+0327}).

\end{enumerate}
We now discuss which {\newtxt of the known} binary pulsars are most suitable for 
this particular test.


\subsection{Figure of merit for detection of SEP violation}
\label{sec:figure_of_merit}

Our simulations indicate that for the parameters affected by SEP violation the 
uncertainty provided by timing is given by:
\begin{eqnarray}
\delta \dot{e} &\simeq& 8.0 \times \frac{\delta t}{x \sqrt{\bar{N} T^3}}\;, \label{eq:edot}\\
\delta \dot{x} &\simeq& 5.3 \times \frac{\delta t}{  \sqrt{\bar{N} T^3}}\;,
\end{eqnarray}
where $\bar{N}$ is the average number of TOAs per unit time (which we assumed to 
be uniform in our simulations), $\delta t$ is the rms of the TOA residuals and 
$T$ is the observing baseline. This also implies, in general, that $\delta 
\dot{e} \simeq 1.5\, \delta \dot{x}/x$. {\newtxt These expressions assume the 
absence of red timing noise. If those effects are present at measurable levels, 
then the improvement with time is slower \cite{kp00}}.

Using the expressions in Section~\ref{sec:SEP_theory} and
the expression for the total mass $M$ derived from
the mass function (eq.~3.15 in \cite{dt92}) we can re-write
$\dot{e}$ as:
\begin{equation}
\dot e = 
  \frac{3}{2} \left( \Delta_{\rm p} - \Delta_c \right)(\hat{\bf{b}} \cdot {\bf{g}})
  \left(\frac{p}{{\cal G} M_c}\right)^{1/2} \;,
\end{equation}
where $ p \equiv x c (1 - e^2) / \sin i$ is
the length of the {\it semi-latus rectum} of the pulsar's orbit and $\hat{\bf{b}}$ is
the unit vector along its length.
With the two last equations we can calculate, in the eventuality
of SEP violation, the {\em significance} of its measurement:
\begin{equation}
\label{eq:significance}
\frac{\dot e}{\delta \dot{e}} \simeq
   1.7 \times 10^{-5} \,{\rm s}^{-3/2} \, 
  \left( \Delta_{p,3} - \Delta_{c,3} \right)(\hat{\bf{b}} \cdot {\bf{g}}_{10})
  \left(\frac{G}{{\cal G}}\frac{T_y^3 \bar{N}_d}{\delta t_\mu^2} \frac{x^3 (1 - e^2)}{m_c \sin i} \right)^{1/2},
\end{equation}
where $m_c \equiv M_c / M_{\odot}$,
${\bf{g}}_{10} \equiv {\bf{g}} / 10^{-10} \rm\, m\, s^{-2}$,
$\Delta_{i,3} \equiv \Delta_{i}/10^{-3}$,
$T_y$ is the timing baseline in years, $\bar{N}_d$
is the number of TOAs per day and $\delta t_\mu$
is the TOA r.m.s.~in $\mu$s. {\newtxt Furthermore, limits on viable 
alternative theories of gravity from pulsar-WD systems \cite{bbv08,fwe+12}
imply that we can assume $G / {\cal G} \simeq 1$,
where $G$ is Newton's gravitational constant}. If there
is SEP violation, then given enough time $T$ it will eventually be detected
to high significance.

\subsection{Current precision of direct SEP violation test}
\label{sec:present_limits}

To evaluate eq.~(\ref{eq:significance}), we must determine
$(m_c \sin i)$ with some degree of precision;
this requires a significant measurement of the Shapiro delay.
{\newtxt Apart from this, $\omega$ and $\Omega$ (the longitude of
periastron and position angle of the line of nodes) are necessary
for determining the absolute orientation of the orbital plane
in space.
Furthermore, the parallax of the system $\pi_x$
is necessary for determining its distance $d$ and its location
in the Galaxy, which are necessary to estimate
the Galactic acceleration ${\bf{g}}$ at that location
and its projection $(\hat{\bf{b}} \cdot {\bf{g}})$
along the direction of the semi-latus rectum}.
With the exception of $\omega$
all of these parameters require high timing precision.
There are only two
systems for which all these parameters have been measured,
PSR~J0437$-$4715 and J1713+0747. The latter has a
significantly larger $x$, which according to eq.~(\ref{eq:significance})
increases its sensitivity to SEP violation. Thus in what follows
we discuss this system in more detail.

Pulsar timing can in principle
be used to detect very low-frequency 
gravitational waves \cite{saz78,det79}.
Presently, several large-scale projects are attempting
to achieve this with
precise, sustained timing of several MSPs
(a {\em pulsar timing array}, or PTA, \cite{vbb+10,fvb+10,dfg+12}).
PSR~J1713+0747 is part of all these efforts because it
is detectable by all major radio telescopes and
is currently one of the three pulsars with the smallest
$\delta t$ known, 0.1$\,\mu$s \cite{dfg+12}.
It has also been precisely timed
for two decades, making its $T$ unusually large.
All of this makes this system extremely sensitive to
$\dot{e}$ caused by SEP violation.
Using a PTA pulsar for this experiment also means that
the SEP violation test proposed here demands no extra time
allocation --- the scientific results can be obtained for
free as an added benefit of ongoing efforts.

We now use TOA simulations to estimate what limits on
$\dot{e}$ can be achieved for this pulsar.
We do this partly because no values and uncertainties
for $\dot{e}$ have been published to date for this pulsar
{\newtxt (an illustration of how unexpected a measurable $\dot{e}$ is,
see Section~\ref{sec:clean})},
but also to estimate future limits on its precision.
In our simulations, we use TOA datasets with
uncertainties, number, start and end times given by Table 1 of
\cite{sns+05} and Table 2 of \cite{dfg+12}; the latter dataset
appears to have started at the end of 2004 and continues to
the present and it completely dominates our simulated
dataset. For the first 6 years 2368 TOAs had been taken,
at an average rate of 1 TOA per day,
we assume in our simulations that 30 TOAs were obtained in a single
session every 30 days; furthermore we assume
that this uniform rate continues at present.

With these assumptions, we obtain $\delta \dot{e} = 1.6 \times 10^{-18}\, \rm 
s^{-1}$ as a realistic uncertainty at the end of 2012. {\newtxt If the
measured value for $\dot{e}$
is consistent with zero, then given} the location
and orbital orientation of this binary, this would result in $\Delta_{\rm p} < 
2\times 10^{-3}$ (95\% C. L.), which is already twice as constraining as the 
current limits from the Damour-Sch\"afer test. If we improve the timing 
precision by a factor of 2 and keep the same timing strategy, then we would have
to wait until 2030 for the precision of the test to increase by one order of 
magnitude. Significantly faster progress might be achievable for MSPs with
larger $x$, or with improvements in timing precision that will be provided by 
the Square Kilometre Array (SKA, \cite{sdl10}).


\subsection{Cleanest binary pulsar experiment}
\label{sec:clean}

Using the results in \cite{dt92}, in particular their eq.~(2.4c), we see that 
the orbital eccentricity $e$, being adimensional, is not affected by the Doppler 
shift $D$ that must necessarily occur during the coordinate transformation from 
the reference frame of the binary to the Solar System Barycenter (SSB).
Another example of such an adimensional quantity is the Shapiro
delay parameter $s$ (or $\varsigma$, \cite{fw10}).
Therefore, it follows that
$e$ is not affected even when there is a {\em change} in that Doppler  
factor, $\dot{D}$\footnote{$D$ and $\dot{D}$ {\newtxt arise from} the first and 
second derivatives of the distance to the system, $d$ \cite{shk70}.}.
At any moment the observed $e$ is the ``intrinsic" 
$e$, plus a small term due to geodetic precession.

It follows from this that if the observed $\dot{e}$ is too small for detection, 
the same limit will also apply to the intrinsic $\dot{e}$, irrespective of 
$\dot{D}$. This is extremely important: as an example, the intrinsic variation
of the orbital period ($\dot{P}_b^{\rm int}$) or the variation of
projected semi-major axis of the orbit ($\dot{x}^{\rm int}$), are
always ``polluted" by $\dot{D}$ \cite{dt91}. and we thus observe
\begin{equation}
\label{eq:doppler_change}
\left(\frac{\dot{P}_b}{P_b}\right)^{\rm obs} =
\left(\frac{\dot{P}_b}{P_b}\right)^{\rm int} - \frac{\dot{D}}{D}, \quad
 \left(\frac{\dot{x}}{x}\right)^{\rm obs} =
\left(\frac{\dot{x}}{x}\right)^{\rm int} - \frac{\dot{D}}{D},
\end{equation}
thus uncertainties in the estimation of $\dot{D}$ impose a
fundamental limit on the precision of the measurement of
$\dot{P}_b^{\rm int}$ and therefore ultimately limit the precision
of radiative tests of gravity \cite{wnt10}.

Another effect that changes $\dot{P}_b^{\rm Int}$ is
mass loss from the binary, which has a lower limit given by
the loss of rotational energy from the pulsar \cite{dt91}; furthermore
outgassing from the companion can in some binaries be so large that no
test of GR can be made. However, such instances of steady mass loss
do not affect the orbital eccentricity \cite{jea25}.

For very wide MSP-WD systems, tidal effects --- in particular
orbital circularization --- are also expected to be extremely small,
since both objects are extremely small compared to the size of the
orbit. They behave effectively as two point masses.

Finally, if GR is the correct theory of gravity, then
{\newtxt $\dot{e}$} is given by \cite{dt92}:
\begin{equation}
\dot{e}^{\rm GR} = \dot{e}^{\rm GW} + \dot{e}^{\rm A},
\end{equation}
where the first term is caused by gravitational wave emission
and the second is caused by a change in the aberration parameter
that is to be expected from geodetic precession.
{\newtxt The leading term of $\dot{e}^{\rm GW}$ is given by \cite{pet64}}:
\begin{equation}\label{eq:edotgw}
  \dot{e}^{\rm GW} = 
    -\frac{304}{15} \, n_b^{8/3} \, 
    \left( T_{\odot} m_c \right)^{5/3} \, \frac{q}{(q + 1)^{1/3}}\, 
    \frac{e ( 1 + 121 e^2/304)}{(1 - e^2)^{5/2}} \,,
\end{equation}
where $q = m_p / m_c$ is the mass ratio and $T_{\odot} \equiv G M_{\odot}/c^3 = 
4.925490947\,\mu$s is one Solar mass in time units. We derive the second term 
from the equations in \cite{dt92}:
\begin{eqnarray}
  \dot{e}^{\rm A} 
    &=& -\frac{1}{\pi} \frac{P}{x} \, n_b^2\, T_{\odot} m_c\, \frac{q + 3/4}{(q + 1)^2}\,
        \frac{e}{(1-e^2)^{3/2}} J(i, \lambda, \eta) \label{eq:edotA} \\
  J(i, \lambda, \eta) 
    &=& \frac{1}{\sin^2 \lambda} \left(\sin i \cos \lambda \sin 2 \eta + 
        \cos i \sin \lambda \cos \eta \right),
\end{eqnarray}
where $\eta$ is the longitude of the projection of the pulsar spin axis in
the plane of the sky measured from ascending node and
$\lambda$ is the angle between the pulsar spin axis
and the line of sight from the pulsar to the Earth (this cannot
be zero, otherwise there would be no pulsations).

For PSR~J1713+0747, these terms are given, respectively, by
$\dot{e}^{\rm GW} = -8.2 \times 10^{-29}\, \rm \,s^{-1}$ 
and $\dot{e}^{\rm A} = 8.2 \times 10^{-28}\, {\rm \, s}^{-1} 
J(i, \lambda, \eta)$.
Since $J(i, \lambda, \eta)$ must be of the order of
unity\footnote{This is a {\newtxt fully} recycled pulsar, therefore
the vast majority of its rotational angular momentum came from
orbiting material. This implies that the spin and orbital angular momenta
must be very closely aligned, therefore $\eta \simeq \pi/2$ and
$J(\lambda, \eta) \simeq 0$ ({\newtxt i.e., much smaller than unity).
This means that for this kind of system} the only likely
contribution to $\dot{e}^{\rm GR}$ comes from $\dot{e}^{\rm GW}$, which is
ten orders of magnitude smaller than {\newtxt the present} $\delta \dot{e}$.},
this means that $\dot{e}^{\rm GR}$ is at least
9 orders of magnitude smaller
than the current $\delta \dot{e}$.
Given the smallness of all polluting terms, we reach the conclusion
that the limits on $\Delta_{\rm p}$ will improve as much as the
experimental precision of $\dot{e}$.

\subsection{Testing SEP with the 1903+0327 binary system.}
\label{sec:1903+0327}

\begin{figure}
  \centerline{\includegraphics[height=8.0cm]{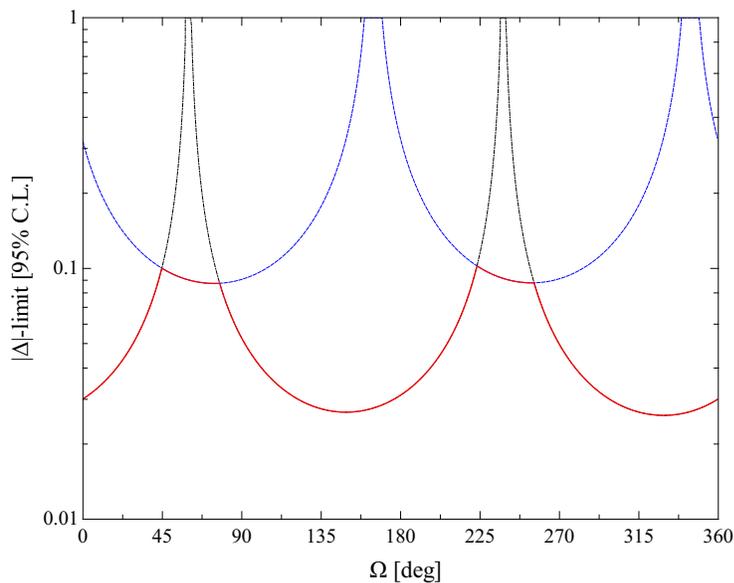}}
  \caption{Limits on $\Delta_{\rm p}$ (95\% C. L.) as a function of 
  the longitude of the ascending node, $\Omega$, for
  the PSR~J1903+0327 binary system, using the values for
  $\dot{x}$ and $\dot{e}$ in \cite{fbw+11} and assuming an orbital
  inclination of $77.47^\circ$ (the other possible inclination is
  $102.53^\circ$, for which we would have similar constraints, but at
  different values of $\Omega$).
  A distance of 6.4 kpc was assumed. The limits in blue are derived
  from the measurement of $\dot{x}$ and the limits in black from
  the measurement of $\dot{e}$; the red line highlights the
  overall upper limit.}
  \label{fig:delta}
\end{figure}

An important advantage of these direct tests of
SEP violation is that we do not need to restrict our study to
binaries with small eccentricities. The binary
system PSR~J1903+0327 \cite{crl+08,fbw+11}
has good timing precision
($\delta t \simeq 1\, \mu\rm s$) and a wide
($x = 105.593\,$lt-s) orbit for which 
$\dot{e} = (14 \pm 6) \times 10^{-17}\,\rm s^{-1}$
and $\dot{x}_{\rm obs} = +21(3) \times 10^{-15}\,\,$lt-s\,s$^{-1}$
have been published \cite{fbw+11}.
A precise distance is not known yet, nor $\Omega$.
However, we can use the parameters
we know precisely (position in the sky, $\sin i$, $\omega$) and a
model of the Galactic potential to estimate
limits on $\Delta_{\rm p}$ for any assumed $d$ and $\Omega$.
These are displayed graphically in Fig.~\ref{fig:delta} for
$d = 6.4 \, \rm kpc$.

Note that we can estimate $\Delta$ limits for {\em every}
value of $\Omega$ because for such eccentric systems
SEP violation also causes a torque in the orbital plane
(eq.~\ref{eq:xdotSEP}). {\newtxt This would add a component
$\dot{x}_{\rm SEP}$ to the kinematic contribution expected for
the system at each particular $\Omega$, $\dot{x}_{\rm kin}$ \cite{ajrt96,kop96}
and the Doppler contribution for each particular $d$ (eq.~\ref{eq:doppler_change}),
such that $\dot{x}_{\rm SEP} = \dot{x}_{\rm obs} - \dot{x}_{\rm kin} + x \dot{D}/D $}.
These limits are not as constraining
as those derived from PSR~J1713+0747, but they are already
interesting because they apply to a massive neutron star
($M_p = 1.667 \pm 0.021\, M_{\odot}$, 99.7\% C.L.), for which no precise
constraints of $\Delta_{\rm p}$ have been obtained until now
(we elaborate on this in Section~\ref{sec:complementarity}).
Furthermore, this limit will increase fast
because a) the $\dot{e}$ and $\dot{x}$ limits were obtained with
only two years of data (from 2008 to 2010), so significant
improvements will be possible in a relatively short timescale and
b) {\newtxt new broadband coherent dedispersion systems will yield a major 
improvement in the timing precision of this system.}

Future infrared interferometric experiments like GRAVITY \cite{bpb+09} will be 
able to measure precisely the astrometric motion of the main sequence
companion to PSR~J1903+0327, allowing therefore a precise, independent 
measurement of $\pi_x$, $i$ and $\Omega$, thus allowing for unambiguous SEP 
tests in this system. {\newtxt Moreover, if indeed SEP is violated at a measurable 
level in PSR~J1903+0327, the eccentric nature of this system allows a unique
cross-check of this since both
$\dot{x}_{\rm SEP}$ and $\dot{e}$ should be influenced in a 
characteristic way: According to eqs.~(\ref{eq:edotSEP}) and 
(\ref{eq:xdotSEP}), the quantity $\dot{x}_{\rm SEP}/(x\dot{e})$ only depends on the orientation and the eccentricity of the pulsar orbit.}


\section{The complementarity of SEP violation and dipolar radiation tests}
\label{sec:complementarity}

As discussed in the Introduction, a theory of gravity which predicts a violation 
of SEP is also expected to predict the emission of dipolar gravitational 
radiation in asymmetric binary systems, like pulsar-WD binaries 
\cite{wil93}. By now there are several binary pulsars that provide tight 
constraints on the existence of dipolar gravitational radiation within 
scalar-tensor gravity, as well as within more general frameworks 
\cite{bbv08,lwj+09,fwe+12,ksm+06}. {\newtxt For these theories, these current 
radiative tests on binary pulsars are} more constraining than SEP tests 
\cite{fwe+12}.

Nevertheless, there are two aspects that support the importance of SEP tests 
with binary pulsars. Firstly, they can be interpreted as generic, direct tests 
for the UFF of strongly self-gravitating bodies, independent of any specific 
gravity theory. Secondly, non-linear strong-field effects, like spontaneous 
scalarization, could be 
limited to very massive neutron stars, {\newtxt which  until now have only been 
observed in wide binary systems like PSR~J1614$-$2230 ($M_{\rm p} \simeq 
1.97\,M_\odot, P_b = 8.7\,$days \cite{dpr+10}) and PSR~J1903+0327 ($M_{\rm p} 
\simeq 1.68\,M_\odot, P_b = 95\,$days \cite{fbw+11}). In these wide systems} a 
dipolar contribution to the gravitational wave damping could, even in future, be 
too small to be detectable, or could not be separable from kinematic effects 
\cite{dt91}, while the deviation from GR could still be measurable in an SEP 
test. In particular since, as outlined in Section~\ref{sec:clean}, the 
$\dot{e}$-test is not ``contaminated'' by external effects.


\section{Future prospects: Globular cluster pulsars, triple systems and mergers}
\label{sec:future}

The external gravitational acceleration by the Galaxy is rather small
($|{\bf{g}}| \sim 2 \times  10^{-10}\,{\rm m\,s^{-1}}$ at the location of the 
Sun). In a stronger external gravitational field {\newtxt SEP violation
would be proportionally more prominent, (eq.~\ref{eq:elfdefs})}. There are 
numerous pulsar binaries known to exist in globular clusters \cite{fre12}, where 
the external acceleration is typically two orders of magnitude larger than in 
the field of the Galaxy \cite{wkm+89,dpf+02,fck+03}. Unfortunately one cannot 
determine the exact location of these systems 
within the globular cluster as one does not have a good handle on the radial 
distance for these pulsars; however this can be somewhat constrained
for systems with negative period derivatives. Furthermore, the
latter provide direct lower limits on ${\bf{g}}$ along the line of sight. If the
semi-latus rectum happened to lie along the line of sight, we would then
automatically have all we need to derive an upper limit on $\Delta_{\rm p}$.

An even stronger external field would arise in a hierarchical triple star 
system, where the pulsar is in a tight orbit with a weakly self-gravitating 
object, and this inner binary falls in the gravitational field of a third 
companion. This would resemble the SEP test done in the Earth-Moon-Sun system, 
but with a strongly self-gravitating object. In the globular cluster M4 there is 
the millisecond pulsar PSR~B1620$-$26 where the inner companion to the pulsar is 
a $\sim 0.3\,M_\odot$ WD and the outer companion is a $\sim$ Jupiter mass 
companion \cite{tacl99,st05}. Unfortunately, the low-mass outer companion yields 
an external acceleration $\bf g$ which is only a factor of a few larger than the 
typical value for the acceleration in the Galactic plane, and therefore does not 
lead to any interesting constraints. The situation would be quite different if 
the outer companion were a $\sim 1\,M_\odot$ star, and even better if it
were also a neutron star. A future detection of a comparable-mass 
triple with a pulsar is not unlikely \cite{trhh08}, particularly since we have
already discovered a binary system, PSR~J1903+0327,
which started its life as a hierarchical triple system and appears to have
become a binary system much later in its evolution
\cite{fbw+11}. Other systems with similar origins might still survive as
hierarchical triples.

Similarly to equation (\ref{eq:DeltaTS}), in a hierarchical triple system one 
would have \cite{de92}:
\begin{equation}
  \Delta_{\rm p} - \Delta_{\rm c} \simeq 
    \alpha_{\rm ex} \, (\alpha_{\rm p} - \alpha_{\rm c}).
\end{equation}
According to this equation, the ideal {\newtxt triple system} combination would be a 
pulsar-WD ($\alpha_{\rm c} = \alpha_0 \ll 1$) or pulsar-black hole 
($\alpha_{\rm c} = 0$) system in the field of a more distant neutron star, for 
which $\alpha_{\rm ex}$, like 
$\alpha_{\rm p}$, could in theory be much larger than $\alpha_0$. Consequently, 
in a hierarchical triple system not only the external gravitational acceleration 
felt by the internal binary would be much larger than for a binary pulsar 
falling in the field of the Galaxy, also the effective scalar coupling of the 
source of the external field could be significantly larger, provided the {\newtxt 
outer companion} is a neutron star.

In such a triple system the challenge lies in obtaining a sufficient number of 
higher order derivatives in the pulsar frequency, in order to constrain the 
orbit and mass of the distant companion such that any effects from the SEP 
violation inflicted on the inner orbit could be separated from ``classical'' 
orbital perturbations \cite{ras94,jr97}. This depends on the details of the 
system, particularly its orbital period and the timing observations (precision, 
time span).

Finally, if the SEP violating interaction is only of limited range, like in the 
massive Brans-Dicke theory of gravity (see \cite{abwz12}), then Galactic binary 
pulsars are insensitive to the corresponding violation of the UFF, since the masses that cause the external gravitational field are at 
large distances. In such a case, direct tests of strong field aspects of the UFF 
would require a hierarchical triple system. 

{\newtxt In about three years from now it is expected that, after their upgrade, the 
ground based gravitational wave detectors will make their first detections of 
gravitational waves. One of the most promising sources for these detectors are 
inspiralling compact binaries, consisting of neutron stars or black holes 
\cite{wal11}. 
This will not only mark the  beginning of the era of gravitational-wave 
astronomy, it will also provide new possibilities to test gravity, including
(indirect) tests of UFF for self-gravitating bodies via its theoretical
connection to the radiative properties of a 
gravity theory. Although, for certain scalar-tensor theories (including JFBD) it 
has been shown that binary pulsar experiments are already more constraining than 
it is expected for the advanced LIGO/VIRGO detectors \cite{wil94,de98,fwe+12}, 
there are many theoretical aspects where ground-based 
gravitational wave antennae will nicely complement binary pulsar experiments in 
the near future, notably theories where the gravitational interaction is partly 
mediated by very short range ($\sim 1$\,lt-s, and less) fields.}


\section*{Acknowledgements}

P.F. gratefully acknowledges the financial support by the European Research
Council for the ERC Starting Grant BEACON under contract no.~279702. {\newtxt We 
also thank Gilles Esposito-Far\`ese for useful comments and stimulating 
discussions} and Nicolas Caballero for a careful review of this paper.
We are grateful to Stanley Deser for bringing \cite{goe12} to our attention.


\end{document}